\documentclass[11pt]{article}

\usepackage{latexsym}
\usepackage{amsfonts}
\usepackage{amssymb}
\usepackage{amsmath}
\usepackage{amsthm}
\newcommand{\bea}{\begin{eqnarray}} 
\newcommand{\eea}{\end{eqnarray}}
\newcommand{\beann}{\begin{eqnarray*}} 
\newcommand{\eeann}{\end{eqnarray*}}
\newcommand{\beq}{\begin{equation}} 
\newcommand{\eeq}{\end{equation}}
\newcommand{\ba}{\begin{array}} 
\newcommand{\ea}{\end{array}}
\newcommand{\ben}{\begin{enumerate}} 
\newcommand{\een}{\end{enumerate}}
\newcommand{\bit}{\begin{itemize}} 
\newcommand{\eit}{\end{itemize}}

\newcommand{\5}{\bar }  
\newcommand{\6}{\partial } 
\newcommand{\7}{\hat }

\newcommand{\sfrac}[2]{\mbox{$\frac{{#1}}{{#2}}$}}
\newcommand{\f}[3]{{f_{#1#2}}^{#3}}
\newcommand{\Ii}{{\mathrm{i}}}

\newcommand{\TR}{\mathit{tr}}

\newcommand{\eff}{{\rm eff}}
\newcommand{\red}{{\rm red}}

\numberwithin{equation}{section}
\newcommand{\bref}[1]{\textbf{\ref{#1}}}

\newcommand{\lieg}{\mathfrak{g}}
\newcommand{\lieh}{\mathfrak{h}}

\def\d{\partial}
\def\defpar{\vartheta}

\def\U{\mathcal U}
\def\g{h}

\def\A{\mathfrak{A}}

\newcommand{\dd}[2]{\frac{\d #1}{\d #2}}

\newcommand{\vddl}[2]{{\frac{\delta #1}{\delta #2}}}

\newcommand{\ab}[2]{\big(#1,\,#2\big)}

\newcommand{\qcommut}[2]{{[#1\stackrel{\star}{,}#2]}}
\newcommand{\commut}[2]{[#1,#2]}

\newcommand{\tr}{{\rm \,Tr}\,}

\begin{document}
\bibliographystyle{../bibtex/utphys}

\title{
\vspace*{-2cm}
%\begin{flushright}
%\normalsize{
%ULB-TH/02-14\\
%MPI-MIS-44/2002\\
%hep-th/0206003
%}
%\end{flushright}
\vspace{1.5cm}
Seiberg-Witten maps and
noncommutative Yang-Mills theories for arbitrary
gauge groups
\vspace*{.5cm}}

\author{Glenn Barnich\thanks{Research 
Associate of the Belgium National Fund for
Scientific Research.}
\vspace{.3cm}
\\
Physique Th\'eorique et Math\'ematique, Universit\'e Libre de
Bruxelles,\\
Campus Plaine C.P. 231, B--1050 Bruxelles, Belgium
\vspace{.3cm}
\and
Friedemann Brandt
\vspace{.3cm}\\
Max-Planck-Institute for Mathematics in the Sciences,\\
Inselstra\ss e 22--26, D--04103 Leipzig, Germany
\vspace{.3cm}
\and
Maxim Grigoriev
\vspace{.3cm}\\
Lebedev Physics Institute,\\
53 Leninisky Prospect,
Moscow 117924, Russia
}
\date{}
\maketitle

\begin{abstract}
Seiberg-Witten maps and
a recently proposed construction of noncommutative Yang-Mills theories 
(with matter fields) for arbitrary gauge groups are
reformulated so that their existence to all orders is manifest.
The ambiguities of the construction which originate from the freedom 
in the Seiberg-Witten map are discussed with regard to the
question whether they can lead to inequivalent models, i.e.,
models not related by field redefinitions. 
\end{abstract}

\newpage

\section{Introduction}

The purpose of this work is to elaborate on 
the construction of
noncommutative Yang-Mills theories for arbitrary gauge groups
proposed in \cite{Jurco:2000ja,Jurco:2001rq} (see
also~\cite{Bonora,Bars:2001iq,Dorn:2002ah} 
for related work).
The idea in \cite{Jurco:2000ja,Jurco:2001rq}
is to use a Seiberg-Witten map for 
building gauge fields and gauge parameters
of the noncommutative theory from Lie algebra valued gauge fields
and gauge parameters of a commutative
gauge theory. The Seiberg-Witten map, which
had been originally established for U(N) theories in \cite{Seiberg:1999vs},
is in \cite{Jurco:2000ja,Jurco:2001rq} applied to the universal
enveloping algebra of the Lie algebra of the gauge group and yields thus 
gauge fields for all elements of the enveloping algebra rather than
only for the Lie algebra itself. 
{}From these enveloping algebra valued gauge fields one
constructs the corresponding field strengths
of the noncommutative theory, and finally the
action in terms of these field strengths.

Our approach is slightly different. As in \cite{Seiberg:1999vs},
the starting point is a noncommutative model
with a Weyl-Moyal star product
but with a suitable finite dimensional space $\U$ of matrices
in place of $u(N)$. 
This model contains unconstrained
``noncommutative'' gauge fields and gauge parameters for a 
basis of $\U$.
By means of a Seiberg-Witten map we express these
gauge fields and gauge parameters in terms of
``commutative'' gauge fields and gauge parameters
that also live in $\U$. 
Finally we set all commutative gauge fields and gauge parameters
to zero except for a subset
corresponding to a Lie subalgebra $\mathfrak{g}$ of $\U$.
In this way one can easily construct noncommutative
gauge theories 
of the same type as in \cite{Jurco:2000ja,Jurco:2001rq}
to all orders in the deformation parameter and for all
choices of $\mathfrak{g}$. 
The inclusion of matter fields is
also straightforward, as we shall demonstrate.

The main difference of our approach from
\cite{Jurco:2000ja,Jurco:2001rq} 
is that we use unconstrained commutative gauge fields and 
gauge parameters for
{\em all} elements of a basis of $\U$ (rather than only of $\mathfrak{g}$)
and drop the unwanted
fields and parameters only at the very end.
An advantage of this approach is
that the existence of the models to all orders 
is manifest. Another advantage is
that it allows one to analyse more systematically
the ambiguities of the construction, especially those which originate from the
freedom in the Seiberg-Witten map pointed out already in 
\cite{Asakawa:1999cu,Jurco:2001rq,Brace:2001fj,Brace:2001rd}.

For completeness, we mention a few references where other approaches
to Seiberg-Witten maps in the Yang-Mills case are discussed. Existence by 
explicit construction has been shown in e.g.\cite{Okuyama:1999ig},
where commutative and  noncommutative versions of Wilson lines were
compared and in \cite{Jurco:2000fs}, where the Seiberg-Witten map was 
computed in
the framework of Kontsevich's approach to deformation quantization. 
An explicit inverse Seiberg-Witten map was given in
\cite{Okawa:2001mv}, where further references are discussed. 

\section{Basic idea}\label{sec:idea}

We work in flat $n$-dimensional spacetime with Minkowski metric. 
In standard (commutative) Yang-Mills theory, the gauge
fields $A_\mu$ take values in a finite-dimensional Lie algebra $\lieh$.
This property is preserved under the gauge transformation
\bea
\label{eq:CGT}
\delta_\lambda A_\mu=D_\mu \lambda=\d_\mu \lambda +\commut{A_\mu}{\lambda}\,,
\eea
for gauge parameters $\lambda$ that are also $\lieh$-valued.
The construction of the gauge invariant Lagrangian
\begin{equation}
L=\frac{1}{2\kappa}\,g_{AB}F^A_{\mu\nu} F^{B \,\mu\nu},
\label{eq:c-Lagrangian}
\end{equation} 
requires in addition the existence of a symmetric bilinear invariant
form $g$ on $\lieh$. The Lagrangian involves the components of 
both $g$ and the curvature 
$F_{\mu\nu}=\d_\mu A_\nu-\d_\nu A_\mu+\commut{A_\mu}{A_\nu}$ in a
basis of $\lieh$.

For noncommutative models, the usual multiplication of functions
is replaced by star-multiplication and a straightforward generalization
of (\ref{eq:CGT}) is
\begin{equation}
\label{eq:NCGT}
\hat \delta_{\hat\lambda} \hat A_\mu=\hat D_\mu \hat\lambda
=\d_\mu \hat \lambda +\qcommut{\hat A_\mu}{\hat \lambda}
\end{equation}
where $\qcommut{\hat A_\mu}{\hat \lambda}$ is the star-commutator. 
To define it, the  fields are assumed to take values\footnote{More
  precisely, we consider the tensor product of $\A$ and the space 
of functions that are formal power series in
the deformation parameter $\defpar$ with coefficients that depend on
$x^\mu$, the fields and a finite number of their derivatives.} in a
finite-dimensional complex associative algebra $\A$ and
$$
\qcommut{f}{g}=f \star g - (-1)^{|f||g|} g \star f
= (f^A*g^B - (-1)^{|f||g|} g^A*f^B) t_A t_B\,,
$$
where we have introduced a basis $t_A$ in $\A$.   
For our purposes it is
sufficient to take as $\A$ a subalgebra of the algebra of constant matrices
with complex entries. We also restrict ourselves to the Weyl-Moyal
star-product, 
\begin{equation}
\label{eq:Moyal}
f \star g = f [\exp(\stackrel{\leftarrow}{\partial_\mu}
\sfrac{\Ii}{2}{\defpar\theta^{\mu\nu}}
\stackrel{\rightarrow}\partial_\nu)]g\,,
\end{equation}
where $\defpar$ is a real deformation parameter and
$\theta$ is a real antisymmetric matrix.

Let us note that the condition that the fields take values in an
associative algebra is sufficient, but not necessary. Indeed, 
in order for the transformation~\eqref{eq:NCGT} to make sense,
it is sufficient that
$\hat A_\mu$ and $\hat \lambda$ take values in some subset $\U
\subset \A$ with the property that the
$\U$ valued fields are closed under star-commutation, i.e., that they
form a star-Lie algebra. These star-Lie algebras can either be 
over $\mathbb{C}$ or over
$\mathbb{R}$. In the complex case, $\U$ has to be 
an associative subalgebra of $\A$, while in the real case, 
it has to be a real subalgebra of $\A$, considered itself as a
real Lie algebra. In the latter case, 
$\U$ is in general neither a vector space over $\mathbb{C}$ (in spite
of the fact that its elements are matrices with complex
entries) nor an associative subalgebra of $\A$.  
The basic examples of real star-Lie algebras are those 
associated to the familiar noncommmutative $U(N)$ models, where
$\U$ is the subspace of antihermitian matrices of
$Mat(N,\mathbb{C})$.

For a given (complex or real) star-Lie algebra, it is
straightforward to construct a Lagrangian which is
invariant up to a total divergence under the
gauge transformations (\ref{eq:NCGT}):
\begin{equation}
\label{L1}
\hat L =\frac{1}{2\kappa}\tr(\hat F_{\mu\nu} \star \hat F^{\mu\nu})
=\frac{1}{2\kappa}\,g_{AB}\hat F_{\mu\nu}^A \star \hat F^{B\,\mu\nu}\,
\end{equation}
where
\bea
\hat F_{\mu\nu}=\d_\mu \7A_\nu-\d_\nu \7A_\mu+\qcommut{\7A_\mu}{\7A_\nu},
\quad g_{AB}=\tr(t_At_B)\,,
\eea
and $\tr$ denotes ordinary matrix trace. Note that the Lagrangian
\eqref{L1} can be also defined for gauge fields
taking values in an (appropriate subspace of an) abstract associative
algebra $\A$ equipped with a trace.
In general, $g_{AB}$ will be degenerate
but this poses no problem in our approach because the noncommutative
model \eqref{L1} is only an intermediate stage in the
construction.

The basic idea to construct a noncommutative deformation for commutative
Yang-Mills models with a given gauge Lie algebra $\lieg$ is the
following: 
\begin{enumerate}
\item  take as $\U$ an appropriate Lie algebra containing $\lieg$
as a subalgebra;

\item reformulate the noncommutative Yang-Mills model
in terms of commutative gauge fields $A_\mu$
and some effective Lagrangian $L^{\eff}$ by using a Seiberg-Witten
map; 

\item reduce consistently the commutative theory described by
  $\U$-valued gauge field $A_\mu$ to the Lie subalgebra $\lieg$ of
  $\U$.
\end{enumerate}

Let us explain these steps in more details, for definiteness
in the (more involved) case where $\lieg$ is a 
Lie algebra over $\mathbb{R}$.  

1. Let $\{t_a\}$ be a basis of a faithful finite
dimensional matrix representation of $\lieg$. Furthermore, suppose that
$g_{ab}=\tr(t_a t_b)$ is an invariant symmetric bilinear and non
degenerate form on $\lieg$. We take $\U$ as an enveloping algebra over
$\mathbb{R}$ of the complexified Lie algebra $\lieg$. More precisely, 
we take $\U$ to be the matrix algebra
generated (over $\mathbb{C}$) by the matrices $t_a$ and complement
the elements $t_a$ by additional elements $t_i$ in such a way that the set
$\{t_A\}=\{t_i,t_a\}$ provides a basis of $\U$ considered
as a vector space over $\mathbb{R}$. 
This implies that $\U$ valued fields form a real 
star-Lie algebra.

2. For the noncommutative theory based on the star-Lie algebra of $\U$
valued fields and 
described by the Lagrangian \eqref{L1}, 
one can construct, as explained in more details
in section \ref{coho}, a Seiberg-Witten map
\begin{equation}
  \begin{split}
    \7A_\mu&=\7A_\mu(\defpar,A,\6A,\6^2A,\dots)=A_\mu+O(\defpar)\,,\\
    \7\lambda&=\7\lambda(\defpar,\lambda,\d\lambda,\ldots,A,\d
    A,\dots)=\lambda+O(\defpar)\,,\\
\end{split}
\end{equation}
required to map the noncommutative gauge transformations \eqref{eq:NCGT}
to the commutative transformations \eqref{eq:CGT}
according to
\begin{equation}
\label{eq:gauge-equiv}
\begin{split}
  \delta_\lambda \7A_\mu(\defpar,A,\6A,\6^2A,\dots)
  &=(\6_\mu\7\lambda+[\7A_\mu\stackrel{\star}{,}\7\lambda])
  (\defpar,\lambda,\6\lambda,\dots,A,\6A,\dots)\,.
\end{split}
\end{equation}
Using such a map, one can reformulate the noncommutative theory described
by the Lagrangian~\eqref{L1} 
and the gauge transformations \eqref{eq:NCGT} 
in terms of the commutative gauge fields
with the gauge transformations~\eqref{eq:CGT} and 
an effective Lagrangian
\begin{equation}
L^\mathrm{eff}[A;\defpar]=\hat
L[\hat A [A;\defpar];\defpar].
\end{equation}
By construction, $L^\mathrm{eff}$ is gauge invariant under 
the commutative gauge transformations up to a total derivative.

3. One can now consistently reduce the model by dropping
all the components of the gauge field and gauge parameter
complementary to the Lie subalgebra $\lieg$ of $\U$.
In the basis $\{t_a,t_i\}$, this means setting to zero the components
$A^i_\mu$ and $\lambda^i$.  Using that $\lieg$
is a Lie subalgebra of $\{t_A\}$, one obtains:
\bea
\left[\delta_\lambda A_\mu\right]_{A^i_\mu=\lambda^i=0}
=\6_\mu\lambda^a t_a+A^b_\mu\lambda^c[t_b,t_c].
\label{eq:GT-reduction}
\eea
On the one hand,
this shows that indeed, it is consistent to set
$A^i_\mu$ and $\lambda^i$ to zero. On the other hand,
it provides the gauge transformations of $A_\mu^a$ in the
reduced theory. We denote these transformations by
$\delta^{\red}_\lambda$. 
\begin{equation}
\label{eq:GT-reduced}
\delta^\mathrm{red}_\lambda A_\mu^a=\d_\mu\lambda^a+A^b_\mu\lambda^c
\f {b}{c}{a},
\end{equation}
where $\f{b}{c}{a}$ are the structure constants
of $\mathfrak{g}$ in the basis $\{t_a\}$,
\bea
[t_b,t_c]=\f{b}{c}{a} t_a.
\eea
The Lagrangian of the reduced theory is given by
\begin{equation}
\label{eq:L-eff-red}
L^{\eff}_{\red}[A^a;\defpar]=
L^{\eff}[A^A;\defpar]\Big|_{A^i=0}\,.
\end{equation}
It is gauge invariant under the gauge
transformations $\delta^{\red}_\lambda$ up to a total
derivative. The resulting model is 
a consistent deformation of a commutative Yang-Mills
theory with gauge algebra $\lieg$, because for $\defpar=0$,
one recovers the standard Yang-Mills Lagrangian.
Even though $L^{\eff}$ is in general complex-valued,
in the case of real $\lieg$ and real $g_{ab}$, the real part
of the Lagrangian $L^{\eff}_{\red}$ is a consistent deformation of the
commutative Yang-Mills theory, because both the real and imaginary
parts of $L^{\eff}_{\red}$ are separately gauge invariant. 

Note that one can also choose to work from the beginning
consistently over $\mathbb{R}$, if the Weyl-Moyal
star-product is taken to be real, i.e., without the imaginary unit
$\Ii$ in formula \eqref{eq:Moyal}. 

Finally, we also note that, in terms of the noncommutative gauge
fields, the effective model described by $L_{\red}^{\eff}$ 
is described by $\hat L$
supplemented by the constraints $A^i_\mu[\hat A^A;\defpar]=0$ involving
the inverse Seiberg-Witten map. This will be
developed in more details in \cite{Bigpaper}. 

\section{Compact gauge groups and matter fields}\label{natural}

Let us now discuss the construction of models which are
based on $u(N)$ valued fields\footnote{We use this term
in a somewhat sloppy way here: in the following ``$u(N)$ valued fields''
live in a space of antihermitian (field dependent) matrices 
which form a star-Lie algebra over $\mathbb{R}$, see below.}.
A physical motivation
for considering such models is that they arise naturally when
one deals with compact Lie algebras $\lieg$. Furthermore, they allow 
one to include directly fermionic matter fields, as we shall
demonstrate below. A mathematical motivation is
that they provide a natural arena for constructing real star-Lie
algebras for the Weyl-Moyal product because the Weyl-Moyal star-commutator of
two antihermitian matrices is again antihermitian \cite{Matsubara}.

So, in the case that the real Lie algebra $\lieg$ admits a faithful
representation by antihermitian $N \times N$ matrices $t_a$, as do the
physically important compact Lie algebras, there is
no need to take as $\U$ the complexified enveloping algebra
of the $t_a$. It is sufficient to take $\U$ equal to $u(N)$, or to
a Lie subalgebra of 
$u(N)$ that contains $\lieg$ as a subalgebra and is
such that $\U$ valued functions form a real star-Lie algebra.
One can then proceed with steps 2 and 3 described in
Section~\bref{sec:idea}, with $u(N)$-valued fields $\7A_\mu=\hat
A^A_\mu t_A$ and $\hat\lambda=\hat\lambda^A t_A$ where all
$\7A^A_\mu$ and $\7\lambda^A$ are real and the $t_A$ are antihermitian. 

To discuss the inclusion of fermionic matter fields,
we introduce a set of Dirac spinor fields which make up
column vectors $\hat\psi$ on which the matrices $t_A$ act, and
noncommutative gauge transformations
\bea
\label{fermiontrafo}
\7\delta_{\7\lambda}\7\psi=-\7\lambda\star \7\psi.
\eea
The Dirac conjugate spinor
fields then transform according to
\bea
\label{fermiontrafo2}
\7\delta_{\7\lambda}\,\5{\!\7\psi}=-\,\5{\!\7\psi}\star\7\lambda^\dagger.
\eea
The straightforward noncommutative
generalization of standard commutative Yang-Mills actions with
fermionic matter fields reads
\begin{equation}
\hat L =\frac{1}{2\kappa}\tr(\hat F_{\mu\nu} \star \hat F^{\mu\nu})
+\Ii \ \5{\!\7\psi}\star\gamma^\mu \7D_\mu\7\psi\,
\label{eq:NC-Lagrangian}
\end{equation}
where, in view of (\ref{fermiontrafo}),
\bea
\7D_\mu\7\psi=\6_\mu\7\psi+\7A_\mu\star\7\psi.
\label{ncaction1}
\eea
Using now (\ref{fermiontrafo2}), one readily verifies that
the Lagrangian (\ref{eq:NC-Lagrangian}) is invariant up to a total 
derivative under the gauge transformations \eqref{eq:NCGT} and
\eqref{fermiontrafo} when $\7\lambda$ is $u(N)$ valued%
\footnote{To include fermionic matter fields
with gauge transformations \eqref{fermiontrafo}
for $\7\lambda$'s
that are not antihermitian, one may double the fermion content
by introducing an additional set of fermions
$\7\chi$ with gauge transformations 
${\hat\delta}_{\hat\lambda }\hat\chi= \hat\lambda^\dagger
\star \hat\chi $. 
The Lagrangian would then contain
$\5{\!\7\chi}\star\gamma^\mu \7D_\mu\7\psi$
in place of $\5{\!\7\psi}\star\gamma^\mu \7D_\mu\7\psi$.
\label{chi}}.

In section \ref{coho}, we shall show that the Seiberg-Witten maps
can be extended such that
\bea
\hat\psi=\hat\psi[\psi,A;\defpar]=\psi+O(\defpar)\,,
\quad
\delta_{\lambda}\7\psi[\psi,A;\defpar]=-\7\lambda[\lambda,A;\defpar]\star 
\7\psi[\psi,A;\defpar]
\eea
where $\delta_{\lambda}$ acts on the commutative fermions
$\psi$ in the standard way:
\bea
\delta_{\lambda}\psi=-\lambda\psi.
\eea
One then proceeds as described in the items 2) and 3) of the previous
section (now with fermions included)   
to arrive at a reduced effective
model with Lagrangian and gauge transformations given by
\bea
&&L^\mathrm{eff}_\mathrm{red}[A,\psi;\defpar]=
\7L[\7A[A;\defpar],\7\psi[\psi,A;\defpar];\defpar]\Big|_{A^i=0}\,,
\\
&&\delta^\mathrm{red}_\lambda A_\mu^a=\d_\mu\lambda^a+A^b_\mu\lambda^c
\f bca,
\quad
\delta^\mathrm{red}_\lambda\psi=-\lambda^at_a\psi.
\eea

\section{Seiberg-Witten maps}\label{coho}

In this section we show the existence of Seiberg-Witten maps
for Yang-Mills models based on arbitrary star-Lie algebra. For
completeness, formulas for fermions are also added. In order
for the them to make sense one should either restrict to
the case of antihermitian gauge fields or add additional fermions as
explained in footnote~\ref{chi}.

The field-antifield formalism 
\cite{Zinn-Justin:1974mc,Zinn-Justin:1989mi,%
Batalin:1981jr,Batalin:1983wj,Batalin:1983jr,Batalin:1984ss,%
Batalin:1985qj} (for reviews, see
e.g. \cite{Henneaux:1992ig,Gomis:1995he})
is an expedient framework
to construct Seiberg-Witten maps 
\cite{Gomis:2000sp,Brandt:2001aa,Barnich:2001mc,Barnich}
because it encodes the action, gauge transformations
and the commutator algebra of the gauge transformations
in a single functional,  the so-called master action which
is a solution to the master equation.
In the language of the field-antifield formalism,
Seiberg-Witten maps are ``anticanonical transformations''
(i.e., transformations of the fields and antifields 
preserving the antibracket)
which turn the master action $\7S$ of a noncommutative model
into the master action $S$ of a corresponding (effective) commutative model,
\bea
\7S[\7\Phi[\Phi,\Phi^*;\defpar],\7\Phi^*[\Phi,\Phi^*;\defpar];\defpar]=
S[\Phi,\Phi^*;\defpar].
\label{SW2}
\eea
Furthermore, the $\7\Phi$'s and $\7\Phi^*$'s are required to agree
with their unhatted counterparts at $\defpar=0$,
\bea
\7\Phi[\Phi,\Phi^*;0]=\Phi,\quad
\7\Phi^*[\Phi,\Phi^*;0]=\Phi^*.
\label{init}
\eea
These maps are generated by a functional $\7\Xi$ satisfying
\bea
\frac{\6\7\Phi}{\6\defpar}=
-\left(\7\Phi,\7\Xi[\7\Phi,\7\Phi^*;\defpar]\right),\quad
\frac{\6\7\Phi^*}{\6\defpar}=
-\left(\7\Phi^*,\7\Xi[\7\Phi,\7\Phi^*;\defpar]\right),
\label{D2}
\eea
where $(\ ,\ )$ is the antibracket for the hatted
fields and antifields.

Explicitly, the master actions are\footnote{For our purposes 
they need not be proper solutions of the master equation
in the strict sense because eventually we shall set $A^i_\mu$ 
and the corresponding ghosts and antifields to zero.},
\bea
\7S[\7\Phi,\7\Phi^*;\defpar]=
\int \left[ 
\7L+\7A^{*\mu}_A(\7D_\mu \7C)^A
+\7\psi^*\7C\star \7\psi+\,\5{\!\7\psi}\star\7C\,\5{\!\7\psi}{}^*
+\7C^*_A(\7C\star\7C)^A\right]d^nx,\quad
\label{ea1}
\\
S[\Phi,\Phi^*;\defpar]=S^\mathrm{eff}[A,\psi;\defpar]+
\int \left[A^{*\mu}_A(D_\mu C)^A
+\psi^* C \psi+\,\5{\!\psi}C\,\5{\!\psi}{}^*
+C^*_A(CC)^A\right]d^nx,
\label{ea2}\eea
where the $\Phi$'s and $\Phi^*$'s and their hatted
counterparts denote collectively the
fields and antifields of the respective formulation
($\{\Phi^M\}\equiv\{A_\mu^A,\psi,C^A\}$ etc.),
the $\7C^A$ and $C^A$ are ghost fields substituting for
the gauge parameters $\7\lambda^A$ and $\lambda^A$, respectively,
$\7L$ is the Lagrangian (\ref{eq:NC-Lagrangian})
and $S^\mathrm{eff}[A,\psi;\defpar]$ will be determined 
in the course of the construction.

A differential equation for these
Seiberg-Witten maps can be obtained by differentiating equation (\ref{SW2})
with respect to $\defpar$. Since the antifield dependent
terms in $S$ do not depend on $\defpar$, the right hand
side of equation (\ref{SW2}) gives just $\6S^\mathrm{eff}[A,\psi;\defpar]/\6\defpar$.
On the left hand side, $\7S$ depends explicitly on $\defpar$
through the star products and implicitly through
$\7\Phi$ and $\7\Phi^*$ because the latter are
to expressed in terms of the $\Phi$ and $\Phi^*$
through the Seiberg-Witten map which also involves $\defpar$.
In this way one obtains, by using \eqref{D2}, the following condition 
on the generating functional $\7\Xi$,
\bea
\frac{\6{\7S[\7z[z;\defpar];\defpar]}}{\6\defpar}
-(\7S[\7z[z;\defpar];\defpar],\7\Xi[\7z[z;\defpar];\defpar])_{z}
=\frac{\6S^\mathrm{eff}[z;\defpar]}{\6\defpar}
\ ,
\label{D3}\eea
where the notation $\7z\equiv (\7\Phi,\7\Phi^*)$, respectively 
$z\equiv (\Phi,\Phi^*)$ has been used.
When expressed in terms of the noncommutative variables, this
condition is equivalent to 
\bea
\frac{\6{\7S[\7z;\defpar]}}{\6\defpar}
=(\frac{\6S^\mathrm{eff}}{\6\defpar})[z[\7z;\defpar];\defpar]
+(\7S[\7z;\defpar],\7\Xi[\7z;\defpar])_{\7z}\ \label{eq47}.
\eea

Explicitly, differentiation of (\ref{ea1}) with respect to
$\defpar$ yields:
\bea
\frac{\6\7S[\7\Phi,\7\Phi^*,\defpar]}{\6\defpar}
= \frac{\Ii}{2}\,\theta^{\alpha\beta}\int d^nx
\left[\frac{2}{\kappa}\,
\TR(\7F^{\mu\nu}\6_\alpha \7A_\mu\star\6_\beta \7A_\nu)
+\Ii \ \5{\!\7\psi}\gamma^\mu \6_\alpha\7A_\mu\star\6_\beta\7\psi
\right.
\nonumber\\
\left.
+\7A^{*\mu}_A\{\6_\alpha \7A_\mu\stackrel{\star}{,}\6_\beta\7C\}^A
+\7\psi^*\6_\alpha\7C\star \6_\beta\7\psi
+\6_\alpha\,\5{\!\7\psi}\star\6_\beta\7C\,\5{\!\7\psi}{}^*
+\7C^*_A(\6_\alpha\7C\star\6_\beta\7C)^A
\right]
\label{D5a}
\eea
The fact that the ghost fields occur in this expression only
differentiated guarantees the existence of a functional
$\7\Xi$ that satisfies (\ref{D3}). This follows from BRST
cohomological arguments that will be given in \cite{Bigpaper}.
It will also be shown there how one obtains
$\7\Xi$ systematically by means of a contracting homotopy.
Here we only give the result. One obtains
\bea
\7\Xi[\7\Phi,\7\Phi^*;\defpar]&=&
\frac{\Ii}{4}\,\theta^{\alpha\beta}\int d^nx
\left[
\7A^{*\mu}_A\{\7A_\alpha\stackrel{\star}{,}\7F_{\beta\mu}+\6_\beta \7A_\mu\}^A
+\7C^*_A\{\7A_\alpha\stackrel{\star}{,}\6_\beta\7C\}^A
\right.
\nonumber\\
&&\left.
+\7\psi^*\7A_\alpha\star(2\6_\beta\7\psi + \7A_\beta\star\7\psi)
+(2\6_\beta \,\5{\!\7\psi}
-\,\5{\!\7\psi}\star\7A_\beta)\star\7A_\alpha\,\5{\!\7\psi}{}^*
\right],
\label{D6}\\
(\frac{\6S^\mathrm{eff}}{\6\defpar})[\Phi[\7\Phi;\defpar];\defpar]&=&
\theta^{\alpha\beta}\int d^nx\left[\frac{\Ii}{\kappa}\,
\TR(\7F^{\mu\nu}\7F_{\alpha\mu}\star\7F_{\beta\nu}
-\frac 14\7F_{\alpha\beta}\7F^{\mu\nu}\star \7F_{\mu\nu})
\right.
\nonumber\\
&&\left.
+\frac 12\ \5{\!\7\psi}\gamma^\mu \7F_{\mu\alpha}\star\7D_\beta\7\psi
+\frac 14\ \5{\!\7\psi}\gamma^\mu\7F_{\alpha\beta}\star\7D_\mu\7\psi
\right].
\label{D7}\eea
Using now (\ref{D6}) in (\ref{D2}) gives the differential equations 
\bea
\frac{\6\7A_\mu}{\6\defpar}&=&
-\frac{\Ii}{4}\,\theta^{\alpha\beta}
\{\7A_\alpha\stackrel{\star}{,}\7F_{\beta\mu}+\6_\beta \7A_\mu\},
\label{hatA}\\
\frac{\6\7\psi}{\6\defpar}&=&
-\frac{\Ii}{4}\,\theta^{\alpha\beta}(2\7A_\alpha\star\6_\beta\7\psi
+\7A_\alpha\star\7A_\beta\star\7\psi),
\label{hatpsi}\\
\frac{\6\7C}{\6\defpar}&=&
-\frac{\Ii}{4}\,\theta^{\alpha\beta}
\{\7A_\alpha\stackrel{\star}{,}\6_\beta\7C\},
\label{hatC}
\eea
which can be integrated to arbitrary order
in $\defpar$ for the initial conditions (\ref{init}).
They generalize the differential equations derived in \cite{Seiberg:1999vs}
for the noncommutative gauge fields and gauge parameters in 
U(N) models to the more general models considered here.
In fact (\ref{hatA}) and (\ref{hatC}) take the same form as
the corresponding equations in the U(N) case. It is also
evident that
(\ref{hatpsi}) reproduces the first order expressions of the Seiberg-Witten
maps for the fermions derived in \cite{Jurco:2000ja,Jurco:2001rq}.
Using a solution to these equations in
(\ref{D7}), the latter provides the commutative effective action
$S^\mathrm{eff}[A,\psi;\defpar]$ in (\ref{ea2}) [notice that
(\ref{D7}) is still expressed in terms of hatted fields]
according to
\bea
S^\mathrm{eff}[A,\psi;\defpar]&=&S_0^\mathrm{eff}[A,\psi]
+\int_0^\defpar d\defpar'\,(\frac{\6S^\mathrm{eff}}{\6\defpar})
[A,\psi;\defpar'],
\label{D8}\\
S_0^\mathrm{eff}[A,\psi]&=&
\int d^nx\,[(2\kappa)^{-1}\TR(F_{\mu\nu}F^{\mu\nu})
+\Ii \ \5{\!\psi}\gamma^\mu (\6_\mu+A_\mu)\psi].
\label{D8a}\eea

\section{Ambiguities of the construction}

In what follows, we need a description of the ambiguity
in the Seiberg-Witten map\footnote{We are grateful
to I.V.~Tyutin for a discussion of the ambiguity in the Seiberg-Witten map
and explaining his unpublished results on that.}. For our purposes
it is sufficient to do this in terms of the variables of the
commutative theory (for simplicity we shall restrict the discussion
to models without matter fields). Given two Seiberg-Witten maps determined
respectively by:
$$
\hat A_\mu=f_\mu[A;\defpar]\,, \quad
\hat\lambda=\g[A,\lambda;\defpar]
$$
and
$$
\hat A^\prime_\mu=f^\prime_\mu[A;\defpar]\,, \quad
\hat\lambda^\prime=\g^\prime[A,\lambda;\defpar]\,,
$$
the composition of one of these maps and the inverse of the other
is a map 
\begin{equation}
f^0_\mu[A;\defpar]=f^{-1}_\mu[f^\prime[A;\defpar];\defpar]\,,\quad
\g^0[A,\lambda;\defpar]=
\g^{-1}[f^\prime[A;\defpar],\g^\prime[A,\lambda;\defpar];\defpar]\,,
\label{cmsw}
\end{equation}
which preserves the gauge structure of commutative theory. More
precisely, it satisfies a commutative counterpart of the gauge equivalence
condition~\eqref{eq:gauge-equiv}
\begin{equation}
\delta_{\lambda}(f^0_\mu[A;\defpar])
=
(\d_\mu \g^0+\commut{f^0_\mu}{\g^0})[A,\lambda;\defpar]\label{cgeq}
\,.
\end{equation}
Hence, any Seiberg-Witten map can be obtained as the
composition of a fixed Seiberg-Witten map and a map preserving the
gauge structure of the commutative theory. 

If, as in the previous section, we introduce in the context of the
antifield formalism a generating functional
$\Xi^0$ for the infinitesimal anticanonical transformation associated to
\eqref{cmsw}, the condition \eqref{cgeq} can be shown \cite{Bigpaper,Barnich}
to be equivalent to 
\begin{equation}
\ab{S}{\Xi^0}=\delta S^\mathrm{eff}[A;\defpar]\,,
\label{eq:CA-def}
\end{equation}
for some antifield independent term $\delta
S^\mathrm{eff}[A;\defpar]$
which is the variation of the effective action under an infinitesimal
field redefinition of $A_\mu$, and a generating functional of the form 
\bea
\Xi^0 = 
\int d^nx
\left(A_A^{*\mu} \sigma^A_\mu+ C^{*}_A\sigma^A\right)
\eea
where $\sigma_\mu=\sigma_\mu[A;\defpar]$ and
$\sigma=\sigma[A,C;\defpar]$. Under some technical assumptions,
the general solution of~\eqref{eq:CA-def} 
has the following form~\cite{Bigpaper}:
\begin{equation}
\sigma_\mu = D_\mu \lambda +
w_\mu\label{eq:C-amb}\,,\qquad\sigma=s\lambda +[C,\lambda] \,. 
\end{equation}
Here, $\lambda=\lambda[A;\defpar]$, $s=(S,\,\cdot\,)$ is the BRST
differential, and $w_\mu$ satisfies
$sw_\mu+\commut{C}{w_\mu}=0$.
The latter implies that $w_\mu=w_\mu[F;\defpar]$ depends on the gauge
potentials and their derivatives only through the 
curvatures $F^A_{\mu\nu}$ and their covariant derivatives.

The finite transformations $A_\mu=f^0_\mu[A^0;\defpar],
\,C=\g^0[A^0,C^0;\defpar]$ associated to $\Xi^0$ are obtained as 
in section~\bref{coho} by solving the differential equations
\begin{equation}
\dd{A_\mu}{\defpar}=-\ab{A_\mu}{\Xi^0}\,,\qquad
\dd{C}{\defpar}=-\ab{C}{\Xi^0}\,,\label{eq:C-diff}
\end{equation}
with boundary condition $A_\mu(0)=A^0_\mu$,
and $C(0)=C^0$. One can then show \cite{Bigpaper}
that the general solution for $f^0_\mu,\g^0$ is given by 
\begin{equation}
\label{eq:st0}
\begin{aligned}
\g^0&=\Lambda_0^{-1} C \Lambda_0 ~+~ \Lambda_0^{-1}s \Lambda_0\,,\\
f^0_\mu&=\Lambda_0^{-1}(A_\mu+W^0_\mu)\Lambda_0~+~ \Lambda_0^{-1}\d_\mu
\Lambda_0\,,
\end{aligned}
\end{equation}
where $\Lambda_0=\Lambda_0[A;\defpar]$ and 
$W^0_\mu=W^0_\mu[F;\defpar]$ satisfies 
$s W^0_\mu+[C,W^0_\mu]=0$.
An arbitrary Seiberg-Witten map $f'_\mu,h'$ 
is given by the composition of
such a map with a fixed Seiberg-Witten map 
$\hat A_\mu=f_\mu[A;\defpar],\hat C = \g[A,C;\defpar]$ 
and can be represented as
\begin{equation}
\label{eq:st}
\begin{aligned}
\g^\prime&=\Lambda^{-1}\star \g \star\Lambda ~+~ \Lambda^{-1}\star
s\Lambda\\
f^\prime_\mu&=\Lambda^{-1}\star( f_\mu+W_\mu)\star 
\Lambda~+~ \Lambda^{-1}\star\d_\mu
\Lambda\,,
\end{aligned}
\end{equation}
with $\Lambda=\Lambda[A;\defpar]$ and 
$W_\mu=W_\mu[A;\defpar]$ satisfying the condition
$s W_\mu +\qcommut{\g}{W_\mu}=0$. This is closely 
related to the form of the ambiguity in the Seiberg-Witten map
discussed in~\cite{Brace:2001rd}.

Let us now discuss the ambiguities in the construction of
the noncommutative Yang-Mills models 
(i.e., in the construction of $L^{\eff}_{\red}$ from
section~\bref{sec:idea}) due to the choice of a particular
Seiberg-Witten
map. First we note that if $w_\mu=0$ then the remaing 
arbitrariness in~\eqref{eq:C-amb} due to $\lambda$ does not 
affect $L^{\eff}_{\red}$. Indeed, the infinitesimal transformation
$A_\mu \to A_\mu + D_\mu \lambda$ is an infinitesimal gauge
transformation which leaves the Lagrangian invariant up to a total
derivative. So a variation of $L^{\eff}_{\red}$ that is not of this
type is necessarily due to a non-vanishing $w_\mu$. The associated 
variation of $L^{\eff}_{\red}$ is 
\begin{equation}
(\delta L^{\eff})_{\red}[A^a;\defpar]=
\left(
\vddl{L^{\eff}}{A^A_\mu}\, w_\mu^A
\right)\Big|_{A^{i}_\mu=0}+\d_\mu k^\mu\,.
\end{equation}
The condition for $L^{\eff}_{\red}+(\delta L^\eff)_\red$
and $L^{\eff}_{\red}$ to determine equivalent theories is the
existence of an infinitesimal field redefinition $A^a_\mu\to
A^a_\mu+V^a_\mu$
such that
\begin{equation}
\left(
\vddl{L^{\eff}}{A^A_\mu}\, w_\mu^A
\right)\Big|_{A^{i}_\mu=0}
=
\vddl{L^{\eff}_{\red}}{A_\mu^a}\,V^a_\mu
+\d_\mu j^\mu\,.
\label{eq:nonUN-cons}
\end{equation}
That is, the infinitesimal variation $(\delta L^{\eff})_{\red}$
must vanish on the equations of motion for $L^{\eff}_{\red}$, up to a
total derivative. In antifield language, this condition   
is equivalent to the existence of a ghost number $-1$ functional
$\Xi_\red$ of the reduced theory such that 
$$
(\delta S^\eff)\Big|_{A^i_\mu=0}=\ab{S_\red}{\Xi_\red}.
$$
We do not have an explicit counterexample, but it seems highly unlikely 
that Eq.~\eqref{eq:nonUN-cons} has a
solution for arbitrary $w_\mu$ satisfying $s
w_\mu+\commut{C}{w_\mu}= 0$. From this
perspective one concludes that the definition
of the model described by $L^\eff_\red$ is in general ambiguous.

Reversing the perspective, \eqref{eq:nonUN-cons} may be used
as a criterion
to distinguish between Seiberg-Witten maps.
One would then consider two Seiberg-Witten maps as equivalent if
their infinitesimal difference is such that
\eqref{eq:nonUN-cons} holds, and otherwise as inequivalent.
It seems to us that this provides a useful
criterion to classify Seiberg-Witten maps used for reduction.

For the construction of $L^\eff_\red$, only
the reduced Seiberg-Witten map $f^\red_\mu=f_\mu|_{A^i_\mu=0}$,
$\g^\red=\g|_{A^i_\mu=0,\,\,C^i=0}$
is relevant. In the case where $\U$ is an enveloping algebra,
$f_\mu^{\red}$ and $\g^{\red}$ reproduce the
enveloping-algebra valued Seiberg-Witten map introduced in
\cite{Jurco:2001rq}. The $so(N)$ and $usp(N)$ noncommutative models 
introduced in \cite{Bonora} fit in the context of section \ref{natural} of the 
present work ($\lieg=so(N)$ respectively $\lieg=usp(N)$ considered
as a subalgebra of $\U=u(N)$ in the fundamental representation). 
The antiautomorphism in \cite{Bonora} (see also 
\cite{Bars:2001iq}) that selects a star-Lie subalgebra of $u(N)$ valued 
functions should in this context be understood as a condition on the 
reduced Seiberg-Witten map.

We also note that the freedom in the Seiberg-Witten map is not
the only source of ambiguities for the construction of
noncommutative Yang-Mills theories of the type considered
here. Another one is the choice of $\U$ or of
the representation matrices $t_A$ respectively.
We have treated the latter 
as initial data for defining a model but, clearly,
there is a lot of freedom in this choice.
Even in the
case without matter fields different
choices can result in inequivalent models,
whereas standard commutative
pure Yang-Mills theory is unique (given $\lieg$)
up to the choice of the bilinear invariant form $g_{ab}$.

\paragraph{Acknowledgements.}
FB thanks Branislav Jur\v{c}o, Julius Wess and especially
Peter Schupp for discussions and explanations of their work.
MG wishes to thank
P.H.~Damgaard, M.~Henneaux, A.M.~Semikhatov, I.Yu.~Tipunin, and especially
I.V.~Tyutin and M.A.~Vasiliev for helpful discussions. He is
grateful to M.~Henneaux for kind hospitality at the Free University of
Brussels where part of this work was done. The work of GB is supported 
in part by the
``Actions de Recherche Concert\'ees'' of the ``Direction de la Recherche
Scientifique-Communaut\'e Fran\c caise de Belgique, by a ``P\^ole
d'Attraction Interuniversitaire'' (Belgium), by IISN-Belgium
(convention 4.4505.86), by Proyectos FONDECYT 1970151 and 7960001
(Chile) and by the European Commission RTN programme HPRN-CT00131, in
which he is associated to K.~U.~Leuven. 
The work of MG was supported
by INTAS grant 00-00262 and RFBR grant 01-01-00906.

%\bibliography{../bibtex/spires}

\providecommand{\href}[2]{#2}\begingroup\raggedright\endgroup

\end{document}